\documentclass[aps,prd,twocolumn,groupedaddress]{revtex4-1}

\usepackage{graphics,graphicx}
\usepackage{amsmath, amssymb}
\usepackage{natbib,hyperref}
\usepackage{mathtools}

\begin{document}

\title{Energy Density Functional of Confined Quarks: an Improved Ansatz}
\author{Udita Shukla}
\author{Pok Man Lo}
\affiliation{Institute of Theoretical Physics, University of Wroclaw,
PL-50204 Wroc\l aw, Poland}
\email{pokman.lo@uwr.edu.pl}

\begin{abstract}
    Density Functional Theory (DFT) is a robust framework for modeling interacting many-body systems, including the equation of state (EoS) of dense matter.  Many models, however, rely on energy functionals based on assumptions that have not been rigorously validated. We critically analyze a commonly used ansatz for confinement, where the energy functional scales with density as  $U \propto n^{\frac{2}{3}}$ . Our findings, derived from a systematic non-local energy functional, reveal that this scaling does not capture the dynamics of confinement. Instead, the energy functional evolves from  $n^2$  at low densities to  $n$  at high densities, governed by an infrared cutoff. These results suggest that models relying on such assumptions should be revisited to ensure more reliable EoS construction.
\end{abstract}

\maketitle

\section{Introduction}

Density Functional Theory (DFT) is a powerful tool for modeling interacting many-body systems~\cite{DFT_bk1, DFT_bk2, DFT_bk3}. While originally developed for electronic structure calculations~\cite{revmodphys.87.897}, its application has extended to diverse fields such as nuclear physics and astrophysics~\cite{Yang:2019fvs}.

The computation of the equation of state (EoS) for strongly interacting dense matter poses significant challenges due to the inherently non-perturbative dynamics of QCD, including confinement and chiral symmetry breaking. Ideally, the energy functional governing the system should be derived from the fundamental Lagrangian, or at least from an effective field-theoretical framework, to ensure a robust connection between the density dependence of macroscopic quantities and the underlying microscopic interactions. However, many approaches rely on ad hoc energy functionals constructed from heuristic assumptions, which often lack justification and fail to capture the essential physics, thereby limiting the reliability and interpretability of their predictions.

In this work, we critically examine an energy density functional proposed in Ref.~\cite{DB01}, which claims to describe confinement through a density scaling of the form $ U \propto n^{{2}/{3}} $. 
This is, however, not supported by our systematic analysis based on a non-local energy functional.
Instead, we demonstrate that the density dependence of the energy functional 
evolves from \( n^2 \) at low densities to linear order \( n \) at high densities, governed by an infrared cutoff. 

\section{Density Functional Theory for Dense Matter}

The thermodynamic potential in the Luttinger-Ward framework is given
by~\cite{Blaizot:2003tw, jean-paul, Leeuwen_2013}:
\begin{equation}
    \label{eq:omega}
    \Omega[S] = -\tilde{{\rm Tr}} \ln \left( S_0^{-1} - \Sigma \right) 
    - \tilde{{\rm Tr}} \left( S \Sigma \right) + \Phi[S],
\end{equation}
where \( S_0 \) is the free propagator, \( S \) is the fully dressed propagator, and \( \Sigma \) is the self-energy. The operator \(\tilde{{\rm Tr}}\) represents integration over momenta, summation over Matsubara frequencies, and a trace over internal degrees of freedom:
\begin{equation}
    \tilde{{\rm Tr}} = \frac{1}{\beta} \sum \int \frac{d^3p}{(2\pi)^3} \, {\rm tr}_D.
\end{equation}

The interaction dynamics are encoded in the \(\Phi\)-functional, which satisfies the variational condition:
\begin{equation}
    \frac{\delta \Phi}{\delta S} = \Sigma.
\end{equation}
Minimizing the thermodynamic potential yields the Dyson equation:
\begin{equation}
    S^{-1} = S_0^{-1} - \Sigma[S].
\end{equation}

In cases where the interaction is approximated by a local potential, the thermodynamic potential simplifies to:
\begin{equation}
    \label{eq:omegaEZ}
    \Omega(n) = -\tilde{{\rm Tr}} \ln \left( S_0^{-1} - \Sigma \right) - \Sigma \, n + U(n),
\end{equation}
where \( n \) is the density. For instance, in the Nambu–Jona-Lasinio (NJL) model~\cite{Klevansky1992,Buballa:2003qv}, the potential is expressed as:
\begin{equation}
    U_{\rm NJL} = -G_{\rm NJL} \, n_S^2,
\end{equation}
with \( G_{\rm NJL} \) being the four-fermion coupling constant and \( n_S \), the scalar density:
\begin{equation}
\label{eq:cond1}
    n_S = \tilde{{\rm Tr}} S.
\end{equation}
The extremization of the thermodynamic potential imposes the self-consistent condition:
\begin{equation}
    \Sigma_{\rm NJL} = \frac{\partial U_{\rm NJL}}{\partial n_S} = -2 G_{\rm NJL} n_S.
\end{equation}
This shows that the NJL model corresponds to a quadratic approximation of the density functional. The local interaction reduces the thermodynamic potential to a function of density rather than a functional.

To incorporate confinement it is necessary to go beyond local interactions. We therefore consider a functional of the momentum-dependent density \(\hat{n}(\vec{p})\):
\begin{equation}
    \label{eq:Ufunc}
    U[\hat{n}] = \int \frac{d^3 p}{(2 \pi)^3} \frac{d^3 q}{(2 \pi)^3} \,
    V(\vec{p}-\vec{q}) \, \hat{n}(\vec{p}) \hat{n}(\vec{q}),
\end{equation}
where \( V(\vec{p}-\vec{q}) \) describes the non-local interactions between pairs of particles.

Connecting a microscopic Lagrangian to an energy density functional of a many-body system is a complex field-theoretical problem without a general solution. However, the form~\eqref{eq:Ufunc} emerges naturally when applying a Hartree-Fock approximation to an effective Lagrangian describing a current-current interaction mediated by a nonlocal, but instantaneous, potential.

The objective is to determine the density dependence of the functional in Eq.~\eqref{eq:Ufunc}. As we will show, the non-local confining potential not only modifies the strength of interactions but also induces an evolving density dependence that varies with density.

In this work, we employ a confining potential:
\begin{equation}
    V(\vec{q}) = \frac{8\pi b}{\vec{q}^{\,4} + \mu_{\rm IR}^4},
    \label{eq:Vq}
\end{equation}
where \( b \) is the string tension and \( \mu_{\rm IR} \) is an infrared regulator. Transforming to configuration space, the potential becomes
\begin{equation}
\begin{split}
    \tilde{V}(\vec{r}) &= b \, \frac{e^{-\frac{\mu_{\rm IR} r}{\sqrt{2}}} \sin\left( \frac{\mu_{\rm IR} r}{\sqrt{2}} \right)}{\frac{\mu_{\rm IR}^2 r}{2}} \\
    &\xrightarrow{\mu_{\rm IR} \to 0} \frac{b \sqrt{2}}{\mu_{\rm IR}} - b r + \mathcal{O}(\mu_{\rm IR}),
\end{split}
\label{eq:Vr}
\end{equation}
thus recovering the expected linear confinement at small \(\mu_{\rm IR}\).  
The divergent constant can be removed by considering a subtracted potential:
\begin{equation}
    \Delta \tilde{V}(r) = \tilde{V}(r) - \tilde{V}(0),
\end{equation}
which eliminates the irrelevant UV divergence at short distances.
The more serious issue arises from the infrared limit (\(\vec{q} \to \vec{0}\)) of the Fourier transform:
\begin{equation}
    \Delta V(\vec{q}) \sim \int_0^\infty dr\, \frac{4\pi r^2}{q r} \, (-b r) \sin(qr),
\end{equation}
which diverges due to the large-\(r\) (IR) contribution.

While the form in~\eqref{eq:Vq} provides a specific IR regularization, one can associate a physical scale \( R_{\rm IR} = 1/\mu_{\rm IR} \) as the distance beyond which the potential saturates. A natural choice for \( R_{\rm IR} \) is the string-breaking scale~\cite{Pennanen:2000yk,Kaczmarek:2005ui}, typically around \(1.3\)~fm, corresponding to \(\mu_{\rm IR} \approx 0.15\)~GeV.

Our goal is to determine the density dependence of the functional in Eq.~\eqref{eq:Ufunc}. As we will show, the nonlocal confining potential not only modifies the strength of interactions but also induces a nontrivial density evolution of the functional.

\section{Improved Ansatz for an Energy Functional}

It has been suggested, notably in Refs.~\cite{DB01,Bastian:2020unt,Ivanytskyi:2022oxv,Bastian:2023hwy}, that an energy functional scaling as 

\begin{equation}
\label{eq:DB_scale}
U \propto n^{\frac{2}{3}} 
\end{equation}
can describe confinement. This scaling is motivated by a dimensional analysis argument~\cite{Ropke:1986qs}:
\begin{equation}
    \frac{\partial U}{\partial n} \sim \Sigma \propto b \, \langle r \rangle \sim b \, n^{-\frac{1}{3}},
\end{equation}
which integrates to yield the proposed scaling:
\begin{equation}
    U \propto n^{\frac{2}{3}}.
\end{equation}

With the non-local energy functional in Eq.~\eqref{eq:Ufunc}, we can test this hypothesis. Specifically, we evaluate the density dependence of the energy functional for a degenerate Fermi gas in its ground state, where fermions occupy states up to the Fermi momentum \( k_F \), i.e., \( \hat{n}(\vec{k}) = \theta(k_F - |\vec{k}|) \). The density is related to \( k_F \) by:
\begin{equation}
    n = \frac{1}{3 \pi^2} k_F^3.
\end{equation}

The energy functional integral~\eqref{eq:Ufunc}, after removing unnecessary prefactors, takes the form:
\begin{equation}
\begin{split}
    \mathcal{J} &= \int_0^{k_F} dp \int_0^{k_F} dq \, \\
    &\frac{1}{2} \, \int_{-1}^{1} dz  \frac{b \, p^2 q^2}{(p^2 + q^2 - 2 p q z)^2 + \mu_{\rm IR}^4}.
\end{split}
\end{equation}
Performing the angular integral over \( z \) analytically yields:
\begin{equation}
    \label{eq:J1}
    \begin{split}
    \mathcal{J} &= \frac{b \, k_F^2}{4 a^2} \int dp_1 dp_2 \, p_1 p_2 \, \times \\
    &\left( \tan^{-1} \frac{(p_1 + p_2)^2}{a^2} - \tan^{-1} \frac{(p_1 -
    p_2)^2}{a^2} \right),
    \end{split}
\end{equation}
where \( p_1, p_2 \) are normalized momenta (\( 0 \leq p_1, p_2 \leq 1 \)), and \( a = \mu_{\rm IR} / k_F \). An important result of this analysis is to show that at $ k_F \ll \mu_{\rm IR} $ (low densities)

\begin{equation}
\label{eq:lowden}
    \mathcal{J} \approx \frac{b \, k_F^6}{9 \mu_{\rm IR}^4} \propto b \, n^2 \,
    \mu_{\rm IR}^{-4},
\end{equation}
and at \( k_F \gg \mu_{\rm IR} \), corresponding to high densities, \( \mathcal{J} \) asymptotes to:

\begin{equation}
\label{eq:highden}
    \mathcal{J} \approx 
    \frac{\pi+4/3}{12} \, \frac{b \, k_F^3}{\mu_{\rm IR}} \propto b \, n \mu_{\rm
    IR}^{-1}.
\end{equation}
Neither the low- nor high-density limits support the \( n^{2/3} \) scaling proposed in Eq.~\eqref{eq:DB_scale}.\footnote{We note, however, that if one sets \(\mu_{\rm IR} = k_F\), the dimensional argument is restored and \( U \sim b\, n^{2/3} \).} Instead, the energy functional exhibits a dynamical change in its density dependence, scaling as \( n^2 \) at low densities and \( n \) at high densities.
A key conclusion is that non-local interactions introduce a characteristic scale that alters the density dependence, which must be accounted for in order to reliably construct the energy functional.

\begin{figure}
	\resizebox{0.50\textwidth}{!}{%
	\includegraphics{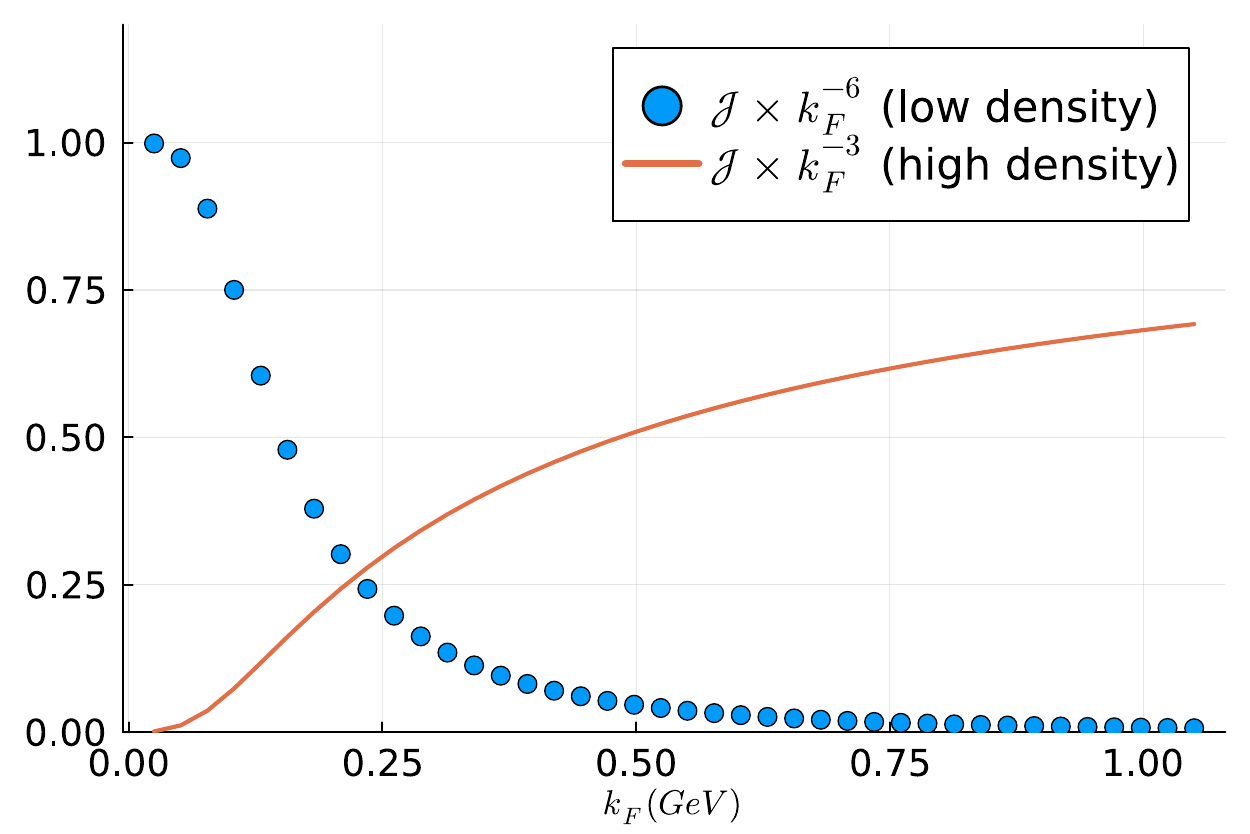}
	}
        \caption{Numerical computation of Eq.~\eqref{eq:J1}, scaled by the low and high density asymptotes (Eq.~\eqref{eq:lowden} and~\eqref{eq:highden}).
        }
\label{fig1}
\end{figure}

In the following, we provide some mathematical details in deriving these key results. To proceed, it is convenient to rewrite the integral~\eqref{eq:J1} in terms of new variables

\begin{equation}
    p_{S, D} = p_1 \pm p_2, 
\end{equation}
such that the integral can be rewritten as

\begin{equation}
    \label{eq:J2}
    \begin{split}
    \mathcal{J} &= \frac{b \, k_F^2}{32 a^2} \, \int_0^2 dp_S \, \int_{d_1}^{d_2}
    dp_D \, (p_S^2-p_D^2) \quad \times \\
    &\left( \tan^{-1} \frac{p_S^2}{a^2} - \tan^{-1}
    \frac{p_D^2}{a^2} \right).
    \end{split}
\end{equation}
The integration limits $d_1, d_2$ are determined from the restriction of $p_1,
p_2$ between 0 and 1, and reads

\begin{equation}
\label{eq:dd}
    \begin{split}
        d_1 &= {\rm Max} \, [-p_S, p_S - 2] \\
        d_2 &= {\rm Min} \, [p_S, 2 - p_S].
    \end{split}
\end{equation}
We first obtain an estimate of $\mathcal{J}$ at large $\mu_{\rm IR}$ (low
densities). In this case we can use the fact that $\tan^{-1} x \approx x$ for
small $x$ to obtain

\begin{equation}
    \begin{split}
    \mathcal{J} &\approx \frac{b \, k_F^2}{32 a^4} \, \int_0^2 dp_S \, \int_{d_1}^{d_2}
    dp_D \, (p_S^2-p_D^2)^2 \\
    &= \frac{b \, k_F^2}{9 a^4} = \frac{b \, k_F^6}{9 \, \mu_{\rm IR}^4}.
    \end{split}
\end{equation}
Thus the density functional scales as $n^2$ as advertised.

At high densities, $\mu_{\rm IR}/k_F \rightarrow 0$. We find that $\mathcal{J}$
diverges as $\mu_{\rm IR}^{-1}$ as dictated by linear confinement. To see this,
we make use of $\tan^{-1} x \approx \frac{\pi}{2} - x^{-1}$ for large $x$ and
approximate the integral as~\footnote{Here we can take $\tan^{-1} \frac{p_S^2}{a^2}
\approx \frac{\pi}{2}$.}

\begin{equation}
    \begin{split}
    \mathcal{J} &\approx \mathcal{J}_{\rm IR} + \Delta \mathcal{J} \\
    \mathcal{J}_{\rm IR} &= \frac{b \, k_F^2}{32 a^2} \, \int_0^2 dp_S \, 
        \int_{-a}^{a} dp_D \, (p_S^2-p_D^2) \, \left( \frac{\pi}{2} -
        \frac{p_D^2}{a^2} \right) \\
    \Delta \mathcal{J} &= \frac{b \, k_F^2}{32 a^2} \, \int_0^2 dp_S \, 
        \int_{d_1^\prime}^{d_2^\prime} dp_D \, (p_S^2-p_D^2) \, \left(
        \frac{a^2}{p_D^2} \right).
    \end{split}
\end{equation}
Here $(d_1^\prime, d_2^\prime)$ are the appropriate integration limits specified by $(d_1, d_2)$ in Eq.~\eqref{eq:dd} but excluding $(-a, a)$. 
The goal is to extract the leading order ($1/a$) contribution from these integrals. A direct calculation yields

\begin{equation}
    \mathcal{J} \approx \frac{b \, k_F^3}{\mu_{\rm IR}} \times \left(
    \frac{\pi-2/3}{12} + \frac{1}{6} \right).
\end{equation}
Thus, the energy functional exhibits a linear scaling with $n$ in the high-density regime. A numerical demonstration of these asymptotic behaviors is presented in Fig.~\ref{fig1}. In this calculation we pick $\mu_{\rm IR} =0.15$ GeV. The convergence to unity indicates that the respective limiting expressions are satisfied. Notably, the high-density limit is attained only at exceedingly large densities, highlighting the slow approach to this regime.

The infrared cutoff  $\mu_{\rm IR}$ can be formally introduced via polarization effects~\cite{cgauge}, effectively accounting for the unquenching the system. This scale can also be dynamically generated, and can vary with temperature and density.

\section{Confinement in Coulomb Gauge QCD}

While the non-local energy functional in Eq.\eqref{eq:Ufunc} provides an improved description of confining quarks, it is not the final solution, and several challenges remain. One significant issue is that the current formulation remains effectively non-relativistic, as Dirac structures are typically neglected, and the self-energy is treated as a scalar. Additionally, there are further model assumptions regarding which types of densities, such as  $n_S, n_V, \dots$ , should be incorporated into the density functional. 

In Refs.~\cite{DB01,Bastian:2020unt,Ivanytskyi:2022oxv}, the energy functional is modeled using only the in-medium part of the scalar density, $n_S^T$, as a variable. This approach neglects the fact that in-medium and vacuum interactions are linked through the same Hamiltonian and should be described within a unified framework.

A related issue stems from equating confinement with infinite quark mass, a common assumption in many studies of dense matter EoS~\cite{DB01,Ivanytskyi:2022oxv}. However, this assumption introduces significant problems. Notably, when the vacuum chiral condensate is computed using Eq.\eqref{eq:cond1}, it fails to yield a finite value, which contradicts the vacuum structure predicted by QCD.

A potential resolution to these issues lies in the confinement mechanism
proposed by QCD in the Coulomb gauge~\cite{cgauge, Alkofer:1989vr,
Reinhardt:2017pyr,Glozman:2007tv}, where an instantaneous potential is explicitly built into the Lagrangian. In this framework, the quark propagator is parameterized as:
\begin{equation}
S^{-1} = (i \omega_n + \mu^\prime_p) \gamma^0 - A_p (\vec{p} \cdot \vec{\gamma}) - B_p I
\end{equation}
which satisfies the gap equations

\begin{equation}
\label{eq:gap}
\begin{split}
    \mu^\prime(\vec{p}) &= \mu + \int \frac{d^3 q}{(2 \pi)^3} \, V(\vec{p}-\vec{q}) \, 
    \frac{1}{2} (\hat{n}_q - \hat{\bar{n}}_q) \\
    A(\vec{p}) &= 1 + \int \frac{d^3 q}{(2 \pi)^3} \, V(\vec{p}-\vec{q}) \, 
    \frac{A_q}{2 \tilde{E}_q} \, \frac{\vec{p} \cdot \vec{q}}{\vec{p}^2}  \, (1 - \hat{n}_q - \hat{\bar{n}}_q) \\
    B(\vec{p}) &= m + \int \frac{d^3 q}{(2 \pi)^3} \, V(\vec{p}-\vec{q}) \, 
    \frac{B_q}{2 \tilde{E}_q} \, (1 - \hat{n}_q - \hat{\bar{n}}_q)\\
    \tilde{E}_q &= \sqrt{A(q)^2 q^2 + B(q)^2}.
\end{split}
\end{equation}
In this case, \( A_p \) and \( B_p \) exhibit infrared divergences, suppressing free quarks, while the chiral condensate remains finite:

\begin{equation}
    n_S = -N_c \int \frac{d^3 q}{(2 \pi)^3} \frac{4 B(q)}{2 \tilde{E}_q},
\end{equation}
as it depends only on the infrared-finite mass function $ M(q) = B(q)/A(q) $. 

The phenomenological implications of the confinement mechanism based on wavefunction renormalization warrant further investigation. Additionally, while the system of equations in Eq.~\eqref{eq:gap} has been explored within the Dyson equation framework, understanding how this translates into the language of DFT presents an important theoretical challenge. We intend to explore these issues in greater depth and report on our findings in a future study.

\section{Going Forward}

Many existing studies \cite{Bastian:2023hwy, Blacker:2023afl, Ivanytskyi:2022wln, Largani:2023oyk, Fischer:2017lag} that attempt to incorporate confinement effects should be critically reevaluated in light of the findings presented here. As noted by others, models that rely heavily on adjustable parameters to reproduce specific results often lack consistency with the principles of QCD~\cite{Jakobus:2022ucs}.

Rather than resorting to parameter fitting and speculative models that obscure the underlying physics, this challenge should be seen as an opportunity to advance the field. The focus must shift toward developing rigorous theoretical frameworks that enable meaningful comparisons with astrophysical observations and heavy-ion collision data, ultimately driving deeper insights into the behavior of dense matter.

Moreover, it is possible that many bulk properties, including the equation of state, may not be sensitive to the intricate details of the underlying interactions. In such cases, further progress may hinge on identifying alternative observables or exploring non-bulk quantities to strengthen the connection to the fundamental dynamics of strong interactions.

\section*{Acknowledgments}
The authors thank Jean-Paul Blaizot, Kenji Fukushima, Owe Philipsen, and Peter Lowdon for inspiring discussions. P. M. L. acknowledges partial support from the Polish National Science Center (NCN) under Opus grant no.
2022/45/B/ST2/01527.

\bibliography{ref}

\end{document}